\documentclass[preprint,showpacs,preprintnumbers,amsmath,amssymb,superscriptaddress,12pt,graphicx]{revtex4}

\usepackage{graphicx}
\usepackage{dcolumn}
\usepackage{bm}
\usepackage{amssymb}

\newcommand{\be}{\begin{equation}}
\newcommand{\ee}{\end{equation}}
\newcommand{\bee}{\begin{eqnarray}}
\newcommand{\een}{\end{eqnarray}}
\newcommand{\ba}{\begin{eqnarray}}
\newcommand{\ea}{\end{eqnarray}}

\begin{document}

\title{Exact Analytic Spectrum of  Relic Gravitational Waves
in Accelerating Universe}

\author{Y. Zhang}
\email{yzh@ustc.edu.cn}\affiliation{ Astrophysics Center, University of Science and Technology of China, Hefei, Anhui, China}
\author{X.Z. Er}
\affiliation{ Astrophysics Center, University of Science and Technology of China, Hefei, Anhui, China}
\author{T.Y. Xia}
\affiliation{ Astrophysics Center, University of Science and Technology of China, Hefei, Anhui, China}
\author{W. Zhao}
\affiliation{ Astrophysics Center, University of Science and Technology of China, Hefei, Anhui, China}
\author{H.X. Miao}
\affiliation{ Astrophysics Center, University of Science and Technology of China, Hefei, Anhui, China}



\begin{abstract}
An exact analytic calculation is presented for the
spectrum of relic gravitational waves in the scenario of accelerating Universe
$\Omega_{\Lambda}+\Omega_m = 1$.
The spectrum formula contains explicitly the parameters of acceleration, inflation,
reheating, and the (tensor/scalar) ratio,
so that it can be employed for a variety of cosmological models.
We find that the spectrum depends on the behavior of the present
accelerating expansion.
The amplitude of gravitational waves for the model $\Omega_{\Lambda}=0.65$ is about
$\sim 50 \%$ greater than that of the model $\Omega_{\Lambda}=0.7$,
an effect accessible to the designed sensitivities of LIGO and LISA.
The spectrum sensitively depends on the inflationary
models with $a(\tau) \propto |\tau|^{1+\beta}$,
and a larger $\beta$ yields  a flatter spectrum, producing  more power.
The current LIGO results rule out
the inflationary models of $\beta \geq -1.8$.
The LIGO with its design sensitivity and the LISA
will also be able to test the model of $\beta=-1.9$.
We also examine the constraints on
the spectral energy density of relic gravitational waves.
Both the LIGO bound  and the nucleosynthesis bound
point out that the model $\beta=-1.8$ is ruled out,
but the model $\beta=-2.0$ is still alive.
The exact analytic results also confirm
the approximate  spectrum and the numerical
one in our previous work.
\end{abstract}



\maketitle


\section{Introduction}

Recently  much progress has been made in
the Laser Interferometer Gravitational waves Observatory
(LIGO) with the typical  sensitivity $10^{-22}$ to $ 10^{-23}$ being reached
in the frequency range  $100\sim 1000$Hz
\cite{abbott} \cite{r-abbott} \cite{b-abbott1} \cite{b-abbott2}.
The chance to detect directly the gravitational waves (GW) has thus increased.
Therefore,  it is necessary to examine the possible objects of detections,
such as the relic GW,
which has a spectrum distributed over a rather broad range of frequencies.
The stochastic background of relic GW
has long been studied
\cite{starobinsky} \cite{rubakov}
\cite{harari}.
The calculations of spectrum generated during the transitions from the inflationary
era  to the radiation-dominated era,  or,  to the matter-dominated era,
have been carried out   \cite{allen} \cite{sahni}
\cite{grishchuk} \cite{maia} \cite{riazuelo} \cite{tashiro} \cite{henriques}.
More recently,
studies has been made on
the effects of the detailed slow-roll inflationary
on the relic GW \cite{gong} \cite{ungarelli}  \cite{smith},
and on the other post-inflationary physical effects
on the relic GW \cite{boyle}.
A constraint on the the tensor-to-scalar ratio $r$ has been derived,
using the CMB-galaxy cross-correlation \cite{cooray}.
The  relic GW can influence CMB and cause magnetic type of CMB polarizations,
which can serve as another distinct signal of the relic GW.
This kind of effects have been studied in Refs.
\cite{Hu,zalda,  kami, keat, zhz}.
On both theoretical and observational issues of the relic GW,
a recent review is given by Grishchuk \cite{r-grishchuk}.

The observations on the SN Ia \cite{riess} \cite{perlmutter} indicate that
the Universe  is currently under accelerating expansion, which may
be driven by the cosmic dark energy ($\Omega_{\Lambda} \sim 0.7$)
plus the dark matter ($\Omega_{m} \sim 0.3$) with
$\Omega_{\Lambda}+\Omega_m=1$ \cite{bahcall} \cite{spergel} \cite{zhy}.
The evolution of relic GW
after being generated during the inflationary stage
depends on the subsequent expansion behaviors of
the  spacetime background.
The  current accelerating expansion  of Universe will have an impact on
the relic GW and  its spectrum.
The spectrum of relic GW has been studied in specific models for
dark energy, such as
the Chaplyngin gas model \cite{fbris} and the X-fluid model \cite{santos}.
In previous study we have studied
the effects on the relic GW  caused by the acceleration of the Universe
for fixed $\Omega_{\Lambda}=0.7$ and $\Omega_{m}=0.3$,
and have obtained an approximate  \cite{zh}, and a numerical spectrum
\cite{zh2} of the relic GW.
It was shown that, in comparison with the decelerating  models,
both the shape and amplitude of the spectrum have been modified
due to the current accelerating expansion.
However,  in the previous work, the dependence of the spectrum upon
the dark energy fraction $\Omega_{\Lambda}$ has not been examined.
Extending these previous studies,
in this paper we present an exact analytic calculation of the spectrum
for any fraction $\Omega_{\Lambda}$ of the dark energy.
We will demonstrate how $\Omega_{\Lambda}$ affects the spectrum,
discuss the dependence of spectrum upon the inflationary models.
We will also examine the resulting spectrum
by comparing with the sensitivity curves of the
gravitational wave detections, such as the LIGO and LISA,
and constrain the corresponding spectral energy density
by the resent LIGO bound and by the nucleosynthesis bound.
The resulting formula of spectrum will contain explicitly the parameter
for the dark energy,
as well as the parameters for the inflationary expansion, the reheating,
the initial normalization of the amplitude, and the ratio of (tensor/scalar),
so that it can be quite general and can be used in other possible applications.
In this way the paper is also to serve as a useful compilation.
Thus we have also listed the main formulae and the relevant specifications
involved in the calculation of the spectrum.
Throughout the paper we adopt
notations similar to that of  \cite{grishchuk}  \cite{zh} for convenience.


\section{ Expansion Stages of the Universe}

The overall expansion of the spatially flat Universe
is described by the  Robertson-Walker metric
$ds^2=a^2(\tau) [  d\tau^2-\delta_{ij}dx^idx^j ]$,
where $\tau$ is the conformal time.
The scalar factor $a(\tau)$ is given by the following
for various stages.

The initial stage (inflationary) \be \label{i} a(\tau ) = l_0 \mid
{\tau} \mid ^{1+\beta},  \,\,\,\,\, -\infty < \tau \leq  \tau_1,
\ee where $1+\beta<0$, and $\tau_1<0$. The special case of
$\beta=-2$ is the de Sitter expansion of inflation.

The  reheating stage
\be
a(\tau) = a_z(\tau-  \tau_p)^{1+\beta_s},
\,\,\,\,\,  \tau_1 \leq  \tau \leq  \tau_s.
\ee
This stage is introduced to allow
a general reheating epoch \cite{grishchuk} \cite{zh}.

The radiation-dominated stage
\be a(\tau) = a_e(\tau  -\tau_e),
\,\,\,\,\, \tau_s \leq  \tau \leq  \tau_2.
\ee

The matter-dominated stage \be a(\tau) =  a_m(\tau  -\tau_m)^2 ,
\,\,\,\,\,  \tau_2  \leq  \tau \leq  \tau_E, \ee where $\tau_E$ is
the time when the dark energy density $\rho_{\Lambda}$ is equal to
the matter energy density $\rho_m$. The  redshift $z_E$ at the
time $\tau_E$ is given by  $1+z_E =
(\frac{\Omega_{\Lambda}}{\Omega_m})^{1/3}$. If the current values
$\Omega_{\Lambda} \sim  0.7$ and $\Omega_m \sim  0.3$ are taken,
then  $1+z_E \sim 1.33$.
For $\Omega_{\Lambda} \sim  0.65$ and
$\Omega_m \sim  0.25$ , then  $1+z_E \sim 1.23$
\cite{zh}.

\begin{figure}
\centerline{\includegraphics[width=10cm]{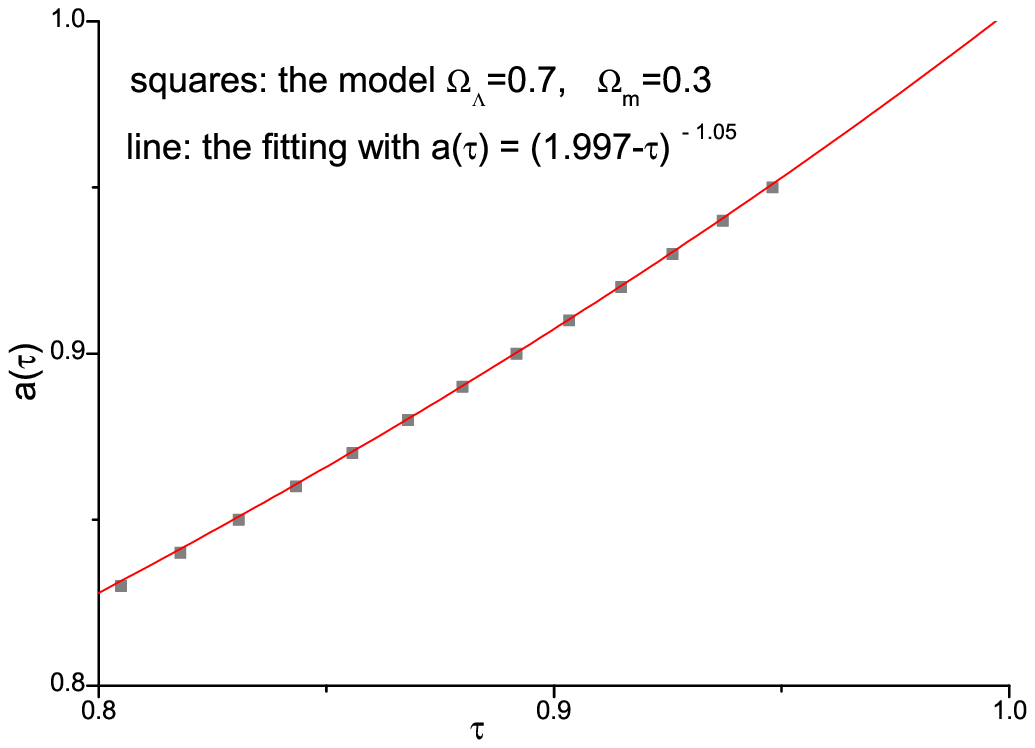}}
\caption{\label{fitting70} For the
accelerating expansion with $\Omega_{\Lambda} = 0.7$
the scale factor $a(\tau)$ can be fitted
by  Eq.(\ref{acc}) with the parameter $\gamma = 1.05$.}
\end{figure}

\begin{figure}
\centerline{\includegraphics[width=10cm]{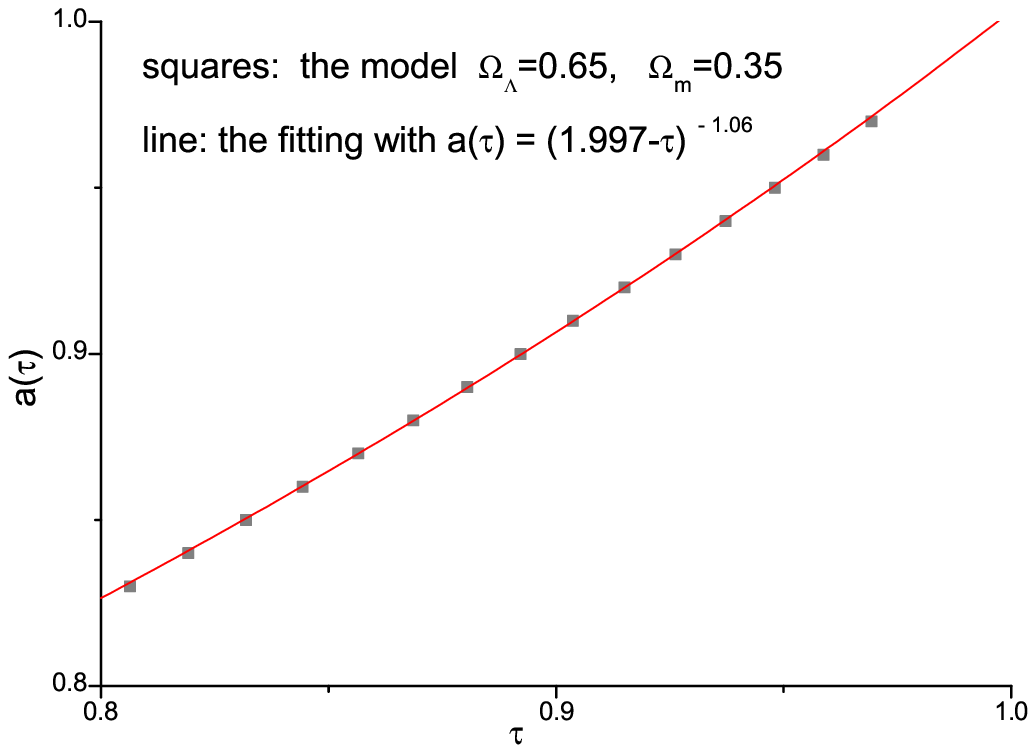}}
\caption{\label{fitting65}For the
accelerating expansion with $\Omega_{\Lambda} = 0.65$
the scale factor $a(\tau)$ can be
fitted by Eq.(\ref{acc}) with  $\gamma = 1.06$.}
\end{figure}

\begin{figure}
\centerline{\includegraphics[width=10cm]{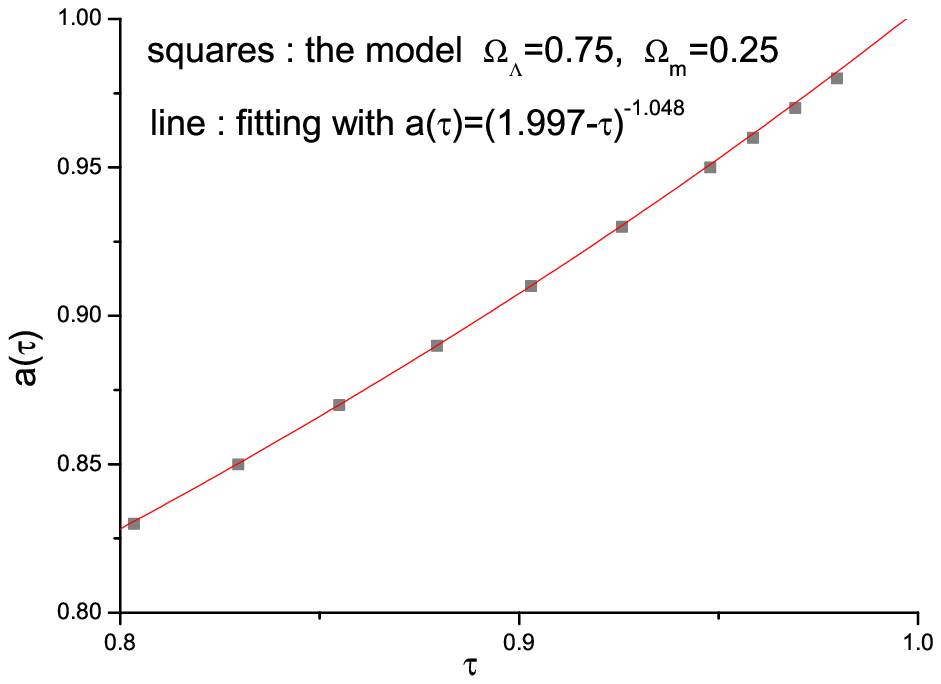}}
\caption{\label{fitting75}For the
accelerating expansion with $\Omega_{\Lambda} = 0.75$
the scale factor $a(\tau)$  can be
fitted by Eq.(\ref{acc}) with $\gamma = 1.048$.}
\end{figure}

The accelerating stage (up to the present time $\tau_H$)
\be \label{acc}
a(\tau) =  l_H |\tau-  \tau_a| ^{-\gamma}, \,\,\,\,\,
 \tau_E \leq  \tau  \leq \tau_H ,
\ee
where the parameter $\gamma=1.0$ is the de Sitter acceleration
for $\Omega_{\Lambda}=  1$ and $\Omega_m=0$.
For  the realistic model with $\Omega_{\Lambda} = 0.7$ and $\Omega_m = 0.3$
at present, we have numerically solved the Friedman equation
\be
(\frac{a'}{a^2})^2= H^2 (\Omega_{\Lambda}+\Omega_m a^{-3})
\ee
where $a'\equiv da(\tau)/d\tau$.
The resulting $a(\tau)$ is plotted in Fig.\ref{fitting70}.
We have found that
the expression of (\ref{acc}) with $\gamma = 1.05$ gives a good
fitting to the numerical solution $a(\tau)$.
Similar calculations show that  $\gamma = 1.06$ fits the model of
$\Omega_{\Lambda} = 0.65$ (in  Fig.\ref{fitting65}),
$\gamma = 1.048$ fits the model $\Omega_{\Lambda} = 0.75$ ( in Fig.\ref{fitting75}),
and $\gamma = 1.042$ fits the model $\Omega_{\Lambda} = 0.80$.
Thus,  for the spatially flat Universe ($\Omega_{\Lambda}+\Omega_m =1$),
as long as the dark energy dominates over the matter component
($\Omega_{\Lambda}>\Omega_m$),
the  generic fitting formula Eq.(\ref{acc}) is effectively valid,
and the range of values for the parameter $\gamma$ are close to $1.0$.
The constant $\tau_a$ in Eq.(\ref{acc}) can be taken to be the same value,
not very sensitive to the various values of
$\Omega_{\Lambda}$ and $\Omega_m$.

There are ten constants in the above expressions of $a(\tau)$,
except  $\beta$, $\beta_s$, and $\gamma$ ,
that are imposed upon as the model parameters.
By the continuity conditions  of $a(\tau)$ and  $a(\tau)'$ at the  four
given joining points $\tau_1$, $\tau_s$, $\tau_2$, and $\tau_E$,
one can fix only eight constants.
The other two  constants can
be fixed by the overall normalization of $a$ and by the observed
Hubble constant as the expansion rate. Specifically,
we put $a(\tau_H)=l_H$ as the normalization, i.e.
\be \label{c}
|\tau_H  -   \tau_a| = 1,
\ee
and the constant $l_H  $ is fixed by the following calculation
\be
\frac{1}{H} \equiv  \left(\frac{a^2}{a'}   \right)_{\tau_H}  =
\frac{l_H}{\gamma}  .
\ee
As we have shown that $\gamma \simeq 1.0$
in the realistic models of acceleration expansion,
so $l_H$ is just the  Hubble radius  at present.
Then everything in the expressions of $a(\tau)$ from
Eq.(1) through Eq.(5) is fixed up.
For  instance, one obtains
\be \label{l0}
  l_0=l_H b\gamma
  \zeta_E^{-(1+\frac{1+\beta}{\gamma})}
  \zeta_2^{\frac{\beta-1}{2}}
  \zeta _s^\beta
  \zeta_1^{\frac{\beta-\beta_s}{1+\beta_s}},
\ee
where $b\equiv |1+\beta|^{-(1+\beta)}$,
$ \zeta_E\equiv  \tau_E/\tau_H$,
$\zeta_2\equiv (\tau_E/\tau_2)^2$,
$\zeta_s\equiv \tau_2/\tau_s$,
and  $ \zeta_1\equiv (\tau_s/\tau_1)^{1+\beta_s}$.

To completely fix the joining conditions
we  need to specify the time instants
 $\tau_{1}$, $\tau_{2}$,  $\tau_s$, and $ \tau_{E}$
 that separate two consecutive expansion stages.
From the consideration of physics of the Universe,
we take the following specifications  \cite{zh}:
$a(\tau_H)/a(\tau_E)=1.33$,
$a(\tau_E)/a(\tau_2)=3454$,
$a(\tau_2)/a(\tau_s)=10^{24}$,
and $a(\tau_s)/a(\tau_1)=300$.
From these, one makes use of the continuity conditions of $a$ and $a'$, and obtains
\[
|\tau_E-\tau_a| =(1+z_E)^{\frac{1}{\gamma}},
\]
\[
|\tau_E-\tau_m| = \frac{2(1+z_E)}{\gamma},
\]
\[
|\tau_2-\tau_m| = \frac{2(1+z_E)}{\gamma\sqrt{3454}},
\]
\[
|\tau_2-\tau_e| = \frac{(1+z_E)}{\gamma\sqrt{3454}},
\]
\[
|\tau_s- \tau_e| = \frac{(1+z_E)\times
10^{-24}}{\gamma\sqrt{3454}},
\]
\[
|\tau_s- \tau_p| = (1+\beta_s) \frac{(1+z_E)\times
10^{-24}}{\gamma\sqrt{3454}},
\]
\[
|\tau_1- \tau_p| = \frac{(1+\beta_s)}{ 300^{\frac{1}{\beta_s+1}} }
         \frac{(1+z_E)\times 10^{-24}}{\gamma\sqrt{3454}},
\]
\be  \label{tau1}
|\tau_1| = \frac{|1+\beta|}{ 300^{\frac{1}{\beta_s+1}} }
         \frac{(1+z_E)\times 10^{-24}}{\gamma\sqrt{3454}}.
\ee
The above expressions all depend on the model parameters $\beta$, $\beta_s$,
and $\gamma$ explicitly, thus depend on $\Omega_{\Lambda}$.
So we can expect that the spectrum of relic GW will
depend on the present acceleration behavior of the Universe
through $\gamma$.

In the expanding Robertson-Walker
spacetime the physical wavelength $ \lambda $ is related to
the comoving wave number $k$ by
\be
 \lambda \equiv \frac{2\pi a(\tau)}{k}.
 \ee
By Eq.(\ref{c}) the wave number corresponding to
the present Hubble radius is
$  k_H = 2\pi a(\tau_H )/l_H  =2\pi $.
There is another wave number,
$ k_E\equiv 2\pi a(\tau_E)H =k_H/(1+z_E)$,
whose corresponding wavelength is the Hubble radius $1/H$ at the time $\tau_E$.


\section{Equation of Gravitational Waves}
Incorporating the perturbations to the
Robertson-Walker metric, one writes
\be
ds^2=a^2(\tau) [  d\tau^2-(\delta_{ij}+h_{ij})dx^idx^j ],
\ee
where $h_{ij}$ is $3\times 3$ symmetric, representing the perturbations.
 The gravitational wave field is
the tensorial portion of $h_{ij}$, which is transverse-traceless
$\partial_i h^{ij}=0$,  $\delta^{ij}h_{ij}=0$,
and the wave equation is
\be
\partial_{\mu}(\sqrt{-g}\partial^{\mu}h_{ij}({\bf{x}} ,\tau))=0 .
\ee
For  a fixed wave vector $\bf k$ and a fixed polarization state
$\sigma$, the wave equation reduces to the second-order ordinary
differential  equation \cite{zh} \cite{grish}
\be \label{h}
 h_k^{(\sigma)''}+2\frac{a'}{a}h_k^{(\sigma)'} +k^2h^{(\sigma)} _k =0,
\ee
 where the prime denotes $d/d\tau$. Since the equation of
 $h_{\bf k}^{(\sigma)} (\tau) $
 for each polarization $\sigma$ is the same, we denote $h_{\bf k}^{(\sigma)}
(\tau) $ by $h_{\bf k}(\tau) $ in the following.
Once the mode function  $h_k(\tau)$ is known,
the spectrum $h(k,\tau)$ of  relic  GW is given by
\be   \label{spectrum}
h(k,\tau) = \frac{4l_{Pl}}{\sqrt{\pi}}k|h_k(\tau)|,
\ee
which is defined by the following equation
\be
\int^{\infty}_0
h^2(k,\tau)\frac{dk}{k}  \equiv <0|h^{ij} ( {\bf x},\tau) h_{ij}(
{\bf x},\tau) |0>,
\ee
where the right-hand-side is the vacuum
expectation value of the  operator $ h^{ij} h_{ij} $.
The spectral  energy density parameter $\Omega_g(k) $ of the
GW is defined through the relation

\[
\frac{\rho_g}{\rho_c}  =  \int \Omega_g(k) \frac{dk}{k},
\]
where $\rho_g = \frac{1}{32\pi G} h_{ij,\, 0} h^{ij}_{\,\, ,\, 0} $ is the
energy density of the GW,
and $\rho_c$ is the critical energy density.
Then, one reads
\be \label{omega}
 \Omega_g(k) = \frac{\pi^2}{3}h^2(k, \tau_H)(\frac{k}{k_H})^2   ,
 \ee
which is dimensionless.
Note that the there might be divergences
 in the integration for $\rho_g$, either infrared or ultraviolet.
As is known, the infrared divergence is avoided if a infrared cutoff is introduced.
This can be done since
the very long waves with wavelengths comparable to, or longer than,
the Hubble length do not contribute to the GW energy  density \cite{zeldovich}.
As for the very short wavelength portion,
the ultraviolet divergences is also avoided
by considering the Parker's adiabatic theorem \cite{parker2},
which states that,  during a transition between expansion epochs with
a characteristic time duration $\Delta t$,
the gravitons created will be suppressed for wavenumbers $k> 1/\Delta t$.
Thus,
the spectrum segments in both the very low and very high frequency ranges
should be discarded from these physical considerations.


\section{  Initial Amplitude of Spectrum}

Regarding to the relic GW,
the initial conditions are taken to be during the inflationary stage.
For a given wave number $k$, the corresponding wave  crossed over the
horizon at a time  $\tau_i$,
i.e. when the wave length was equal to the Hubble radius:
$\lambda_i = 2\pi a(\tau_i)/k$  to $1/H(\tau_i)$.
From Eq.(\ref{i}) yields $H(\tau_i) = l_0^{-1}|1+\beta|\cdot |\tau_i|^{2+\beta}$,
and,  for the case of exact de Sitter expansion
of $\beta =-2$,  one has $H(\tau_i)=l_0^{-1}$.
Thus  a different $k$ corresponds to a different time $\tau_i$. Now choose
the initial condition of the mode function $h_k(\tau)$ as
\be
|h_k(\tau_i)| = \frac{1}{a(\tau_i)}.
\ee
Then the initial amplitude of the  spectrum is \cite{grishchuk} \cite{zh}
\be \label{inv}
h(k, \tau_i) = A(\frac{k}{k_H})^{2+\beta} ,
\ee
where the constant
\be \label{A}
A=8\sqrt{\pi}b \frac{l_{Pl}}{l_0}.
\ee
The power spectrum for the primordial perturbations of  energy density
is $P(k)\propto |h( k, \tau_H)|^2 $, and its spectral index $n$ is defined as
$P(k) \propto  k^{n-1}$.
Thus one reads off the relation $n = 2\beta +5$.
The exact de Sitter  expansion of $\beta = -2$ leads to $n=1$,
yielding an initial spectrum independent of $k$,
called the scale-invariant primordial spectrum.
Other values of $\beta$ will differ from the scale-invariant one.

As is known,
any calculation of the spectrum of the relic GW always has some
overall uncertainty, originating from the normalization
of the amplitude.
Currently, from the observational perspective,
the best that one can do is to use the CMB anisotropies
to constrain the amplitude, as they receive the contributions from
both the  scalar (density) and the tensorial (GW) primordial perturbations.
However,  there is a well known problem of how much relative contribution
is from the relic GW, in comparison with
the scalar type contribution (the density perturbations).
There have been a number of discussion on the ratio of the relic GW to
the scalar contribution,
\be\label{ratio}
r= P_h/P_s.
\ee
Theoretically, it is, in our view, a problem of initial conditions
on the ratio of the scalar and tensorial modes of comic perturbations.
So far, in regards to the very long wavelength,
some preliminary conclusion on the upper limit of GW contributions
has been given,  based  upon the analysis on WMAP
and the observational results of SDSS, for instance,
$r< 0.37$  (95\% c.l.) \cite{Peiris} \cite{Seljak}.
The final conclusion on this issue might be eventually
rely on the more observations of CMB anisotropies
and polarization (such as the Planck project in near future).
In the following, the ratio $r$ is treated as a parameter,
representing the relative contribution
by the relic GW to the CMB anisotropies $\Delta T/T$ at low multipoles.
This will  determine the overall factor $A$  in (\ref{inv}).
Using the observed CMB anisotropies \cite{spergel}
is $\Delta T/T \simeq 0.37 \times 10^{-5}$ at $l \sim 2$,
which corresponds to the anisotropies on the scale of Hubble radius, we put
\be \label{initialcd}
h(k_H,
\tau_H) =  0.37  \times 10^{-5}r.
\ee
Then the spectrum
$h(k, \tau_H)$ at the present time $\tau_H$ is fixed.
If we take the upper limit $r = 0.37$,
then $h(k_H, \tau_H) \simeq 0.14 \times10^{-5}$.
For smaller $r$,
our  calculation is still similar except the resulting spectrum is reduced by
the corresponding numerical factor.


\section{Analytic Solution}

Writing the mode function $h_k(\tau)=\mu_k(\tau)/a(\tau)$  in Eq.(\ref{h}),
 the equation for $\mu_k(\tau)$  becomes
 \be \label{mu}
\mu''_k+(k^2-\frac{a''}{a})\mu_k=0.
\ee
 For a scale factor of  power-law form
 $  a(\tau )   \propto \tau^{\alpha}$,
 the general exact solution is of the following form
 \[
\mu_k(\tau) = c_1(k\tau)^{\frac{1}{2}}J_{\alpha-\frac{1}{2}}(k\tau)
+ c_2(k\tau) ^{\frac{1}{2}}J_{\frac{1}{2}-\alpha }(k\tau),
 \]
 where the constant $c_1$ and $c_2$ are to determined by
continuity of the function $\mu_k(\tau)$ and the  time
derivative $(\mu_k(\tau)/a(\tau))'$ at the time instance
joining two consecutive  stages.

 The inflationary stage has the solution
 \be
 \mu_k(\tau)= x^{\frac{1}{2}}
 [A_1J_{\beta+\frac{1}{2}}(x)+A_2J_{-(\beta+\frac{1}{2})}(x)],
 \,\,\,\, -\infty < \tau \leq  \tau_1,
 \ee
where  $x\equiv k\tau$, and the two constants
$A_1$ and $A_2$, determining the initial states,  are  taken to be
 \be \label{a}
 A_1=-\frac{i}{\cos\beta\pi}\sqrt{\frac{\pi}{2}}e^{i\pi\beta/2},
\,\,\,\,\,\,\,
 A_2=iA_1e^{-i\pi\beta},
 \ee
both are independent of k.
With Eq.(\ref{a}) the mode function $\mu_k(\tau)$
is proportional to the  Hankel's  function $H_{\beta+\frac{1}{2}}^{(2)}$,
\be
\mu_k(\tau) =  A_1 e^{-i\pi\beta}\sin(\beta\pi+\frac{\pi}{2})
x^{\frac{1}{2}} H_{\beta+\frac{1}{2}}^{(2)}(x),
\ee
which, in the high frequency limit,  approaches to the positive frequency mode
\[
\lim_{k \rightarrow \infty} \mu_k(\tau) \rightarrow e^{-ik\tau}.
\]
Thus  the initial state fixed by Eq.(\ref{a}) corresponds to
the so-called adiabatic vacuum in the high frequency limit
\cite{parker} \cite{bunch}.

The reheating stage has
\be
\mu_k(\tau)=t^{\frac{1}{2}}
[ B_1J_{\beta_s +\frac{1}{2}}(t)+ B_2J_{-\beta_s-\frac{1}{2}}(t)],
\,\,\,\,\,\, \tau_1< \tau \leq \tau_s,
\ee
where the variable $t\equiv k(\tau-\tau_p)$,
and the two coefficients  $B_1$ and $B_2$
are fixed by joining the functions
$\mu_k(\tau)$ and $(\mu_k(\tau)/a(\tau))'$ continuously
at the time $\tau_1$ when the reheating epoch begins:
\bee \label{d1}
B_1  =
 \sqrt{\frac{x_1}{t_1}}
\frac{J_{\beta+\frac{1}{2}}(x_1)J_{-\beta_s-\frac{3}{2}}(t_1)
+J_{\beta+\frac{3}{2}}(x_1)J_{-\beta_s-\frac{1}{2}}(t_1)}
{J_{\beta_s+\frac{1}{2}}(t_1)J_{-\beta_s-\frac{3}{2}}(t_1)
+J_{-\beta_s-\frac{1}{2}}(t_1)J_{\beta_s+\frac{3}{2}}(t_1)}A_1
       \nonumber \\
+\sqrt{\frac{x_1}{t_1}}
\frac{J_{-\beta-\frac{1}{2}}(x_1)J_{-\beta_s-\frac{3}{2}}(t_1)
-J_{-\beta-\frac{3}{2}}(x_1)J_{-\beta_s-\frac{1}{2}}(t_1)}
   {J_{\beta_s+\frac{1}{2}}(t_1)J_{-\beta_s-\frac{3}{2}}(t_1)
   +J_{-\beta_s-\frac{1}{2}}(t_1)J_{\beta_s+\frac{3}{2}}(t_1)}A_2,
\een
\bee  \label{d2}
B_2  =
\sqrt{\frac{x_1}{t_1}}
\frac{J_{\beta+\frac{1}{2}}(x_1)J_{\beta_s+\frac{3}{2}}(t_1)
-J_{\beta+\frac{3}{2}}(x_1)J_{\beta_s+\frac{1}{2}}(t_1)}
 {J_{\beta_s+\frac{1}{2}}(t_1)J_{-\beta_s-\frac{3}{2}}(t_1)
 +J_{-\beta_s-\frac{1}{2}}(t_1)J_{\beta_s+\frac{3}{2}}(t_1)}A_1
  \nonumber \\
+\sqrt{\frac{x_1}{t_1}}
\frac{J_{-\beta-\frac{3}{2}}(x_1)J_{\beta_s+\frac{1}{2}}(t_1)
+J_{-\beta-\frac{1}{2}}(x_1)J_{\beta_s+\frac{3}{2}}(t_1)}
     {J_{\beta_s+\frac{1}{2}}(t_1)J_{-\beta_s-\frac{3}{2}}(t_1)
     +J_{-\beta_s-\frac{1}{2}}(t_1)J_{\beta_s+\frac{3}{2}}(t_1)}A_2
\een
with $x_1\equiv k\tau_1$, $t_1 \equiv k(\tau_1-\tau_p)$,
and $(1+\beta_s)x_1=(1+\beta)t_1$,
which follows from the continuity of $a(\tau)$ and $a'(\tau)$ at the time $\tau_1$.

The radiation-dominated stage has
\be
\mu_k(\tau)= C_1e^{-iy}+ C_2e^{iy},
\,\,\,\,\,\,\,\,   \tau_s \leq  \tau \leq  \tau_2,
\ee
where the variable $ y \equiv  k(\tau-\tau_e) $,
and  $C_1$ and $C_2$  are  given by
\be \label{b1}
C_1=\frac{e^{iy_s}t_s^{\frac{1}{2}}}{2i}
\left\{[(i-\frac{1}{y_s})J_{\beta_s+\frac{1}{2}}(t_s)
+J_{\beta_s +\frac{3}{2}}(t_s)] B_1+
[(i-\frac{1}{y_s})J_{-\beta_s -\frac{1}{2}}(t_s)
-J_{-\beta_s -\frac{3}{2}}(t_s)] B_2\right\},
\ee
\be \label{b2}
C_2=\frac{-e^{-iy_s}t_s^{\frac{1}{2}}}{2i}
\left\{[-(i+\frac{1}{y_s})J_{\beta_s +\frac{1}{2}}(t_s)
+J_{\beta_s +\frac{3}{2}}(t_s)] B_1+
[-(i+\frac{1}{y_s})J_{-\beta_s -\frac{1}{2}}(t_s)
-J_{-\beta_s -\frac{3}{2}}(t_s)] B_2\right\},
\ee
where $t_s \equiv k (\tau_s-\tau_p)$, $y_s \equiv k(\tau_s-\tau_e)$,
and $t_s=(1+\beta_s)y_s$.

The  matter-dominated  stage has
\be
\mu_k(\tau)=\sqrt{\frac{\pi
z}{2}}[D_1J_{\frac{3}{2}}(z)+ D_2J_{-\frac{3}{2}}(z)],
\,\,\,\,\, \tau_2  \leq  \tau \leq  \tau_E ,
\ee
where $z\equiv k(\tau-\tau_m)$, and  $D_1$ and $D_2$  are given by
\be  \label{c1}
D_1=[-e^{iy_2}-\frac{i}{2y_2}e^{iy_2}+\frac{e^{iy_2}+e^{-3iy_2}}{8y_2^2}]C_1+
    [-e^{-iy_2}+\frac{i}{2y_2}e^{-iy_2}+\frac{e^{-iy_2}+e^{3iy_2}}{8y^2_2}]C_2,
\ee
\be \label{c2}
D_2=[ie^{iy_2}-\frac{e^{iy_2}}{2y_2}-\frac{i}{8y_2^2}(e^{iy_2}-e^{-3iy_2})]C_1-
    [ie^{-iy_2}+\frac{e^{-iy_2}}{2y_2}+\frac{i}{8y_2^2}(e^{3iy_2}-e^{-iy_2})]C_2,
\ee
with $y_2 \equiv k(\tau_2-\tau_e)$.

The accelerating  stage has
\be \label{m}
\mu_k(\tau)=\sqrt{\frac{\pi s}{2}}[E_1J_{\gamma +\frac{1}{2}}(s)
+E_2J_{-\gamma -\frac{1}{2}}(s)], \,\,\,\,\,  \tau_E \leq  \tau
\leq \tau_H ,
\ee
where  $ s\equiv k(\tau-\tau_a)$, and  $E_1$ and
$E_2$  are given by
\bee \label{e1}
 E_1= \Delta^{-1} \frac{z_{E}}{s_{E}}
 \left\{
 J_{\frac{3}{2}}(z_{E})[-\frac{J_{-\gamma-\frac{1}{2}}(s_E)}{s_E}
 -J_{-\gamma-\frac{3}{2}}(s_E)]
 -J_{\frac{5}{2}}(z_{E})J_{-\gamma-\frac{1}{2}}(s_E)
\right\}D_1
\nonumber  \\
+ \left\{
  J_{-\frac{3}{2}}(z_{E})[-\frac{J_{-\gamma-\frac{1}{2}}(s_E)}{s_E}
  -J_{-\gamma-\frac{3}{2}}(s_E)]
 +J_{-\frac{5}{2}}(z_{E})J_{-\gamma-\frac{1}{2}}(s_E)
 \right\}D_2,
\een \bee  \label{e2}
 E_2= \Delta^{-1} \frac{z_{E}}{s_{E}}
 \left\{
 J_{\frac{5}{2}}(z_{E})J_{\gamma+\frac{1}{2}}(s_E)-J_{\frac{3}{2}}[
 -\frac{J_{\gamma+\frac{1}{2}}}{s_E}(s_E)
 +J_{\gamma+\frac{3}{2}}(s_E)]
\right\}D_1
\nonumber \\
+
 \left\{
 -J_{-\frac{5}{2}}(z_{E})J_{\gamma+\frac{1}{2}}(s_E)
 -J_{-\frac{3}{2}}[-\frac{J_{\gamma+\frac{1}{2}}}{s_E}(s_E)
 +J_{\gamma+\frac{3}{2}}(s_E)]
 \right\}D_2.
\een
\be
 \Delta =J_{\gamma+\frac{1}{2}}(s_E)[
 -\frac{J_{-\gamma-\frac{1}{2}}(s_E)}{s_E}-J_{-\gamma-\frac{3}{2}}(s_E)]
 -J_{-\gamma-\frac{1}{2}}(s_E)[-\frac{J_{\gamma+\frac{1}{2}}(s_E)}{s_E}
 +J_{\gamma+\frac{3}{2}}(s_E)]
 \ee
where $z_E \equiv k(\tau_E-\tau_m)$,  $s_E \equiv
k(\tau_E-\tau_a)$, and $ \gamma z_E=-2s_E$.

With all these coefficients having been fixed,
the mode function $h_k(\tau_H)$ is known
as a function of the wave number $k$  at present time $\tau_H$,
so is the  spectrum
\be\label{spect}
h(k,\tau_H)= \frac{4l_{Pl}}{\sqrt{\pi}}k|h_k(\tau_H)|,
\ee
as defined in Eq.(\ref{spectrum}).
The above results form a useful compilation for computing the relic GW.
To make use of the formulation (\ref{spect}),
one substitutes  $h_k(\tau_H) = \mu_k(\tau_H)/a(\tau_H)$,
where $\mu_k(\tau_H)$ is given in Eq.(\ref{m}).
Of course, to specify $\mu_k(\tau_H)$,
all the coefficients $E_1$ , $E_2$ throughout $A_1$, $A_2$
have to be employed.
One may, in his own computation,
choose proper values of the parameters $\beta$, $\beta_s$, and $\gamma$
for the specific expansion behavior,
as well as the initial amplitude $A$ in Eq.(\ref{initialcd}).

\begin{figure}
\centerline{\includegraphics[width=10cm]{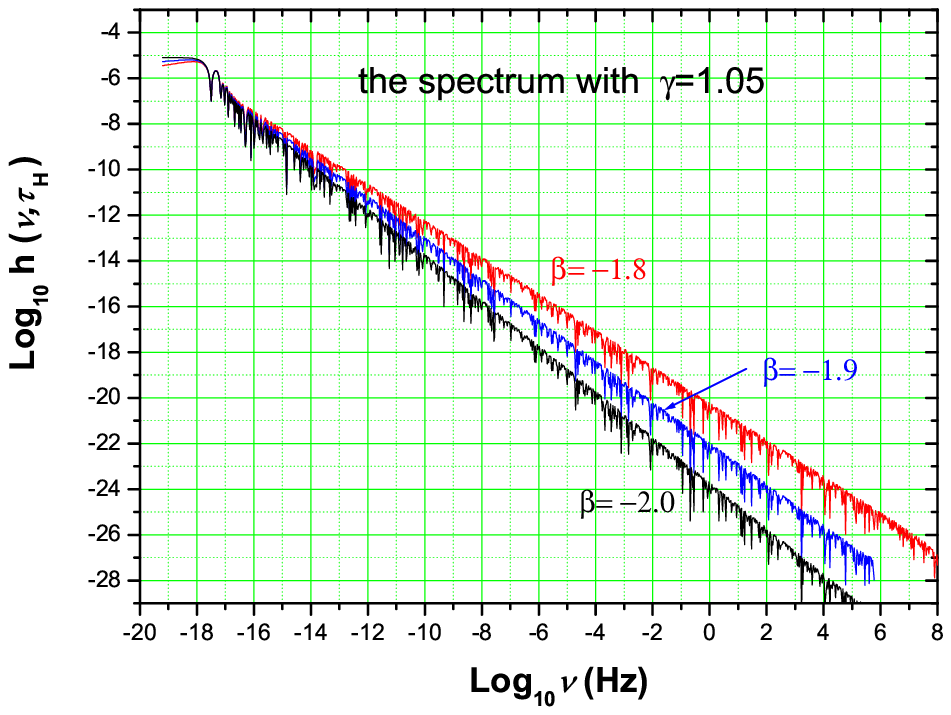}}
\caption{\label{amplitude-105}For a fixed acceleration parameter
$\gamma=1.05$ the exact spectrum $h(\nu,\tau_H)$ is plotted for
three inflationary models of $\beta=-1.8$, $\beta=-1.9$, and
$\beta=-2.0$, respectively.}
\end{figure}

\begin{figure}
\centerline{\includegraphics[width=10cm]{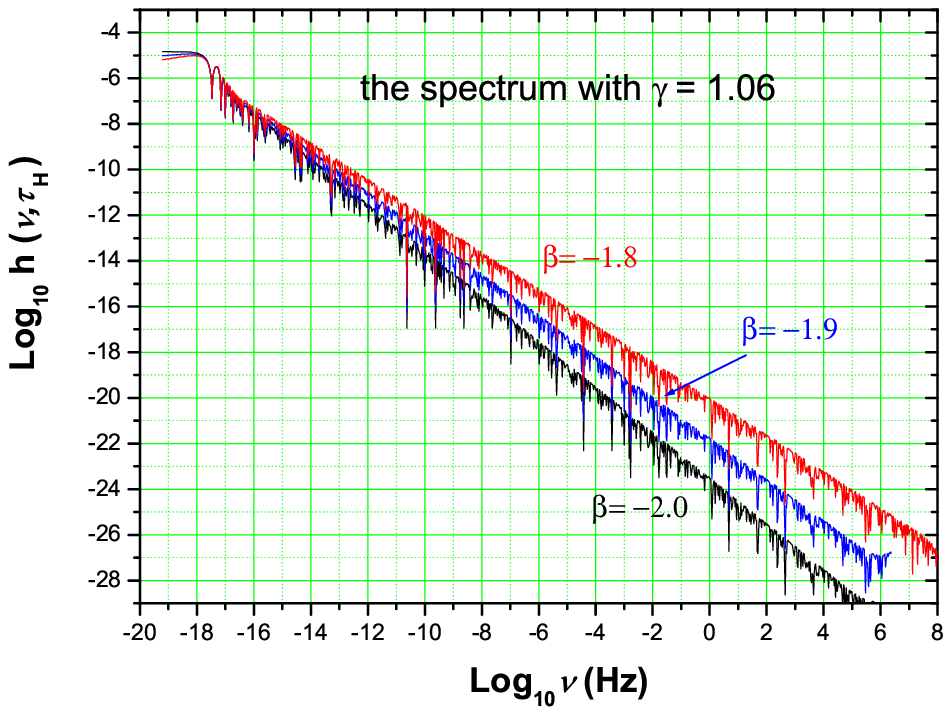}}
\caption{\label{amplitude-106}For a fixed acceleration parameter
$\gamma=1.06$ the exact spectrum $h(\nu,\tau_H)$ is plotted for
three inflationary models of $\beta=-1.8$, $\beta=-1.9$, and
$\beta=-2.0$, respectively.}
\end{figure}

For illustrations,
taking the (tensor/scalr) ratio in Eq.(\ref{ratio}) $r=0.37$,
we have plotted the exact spectrum $h(k,\tau_H)$
as a function of the frequency $\nu=k/2\pi a $ in
Fig.\ref{amplitude-105} for $\gamma=1.05$,
and in Fig.\ref{amplitude-106} for $\gamma=1.06$.
In each of these figures of fixed $\gamma$,
three spectra are shown for three inflationary models with
$\beta=-1.8, -1.9$, and $ -2.0$,
and the parameter $\beta_s =  0.598$,    $ -0.552$, and $-0.689$
are taken, respectively \cite{zh}.
As these figures show,
the spectrum is scale-invariant
with a flat segment in the range $\nu \leq 10^{-18}$Hz
and a slope segment in the range $\nu \geq 10^{-18}$Hz.

\begin{figure}
\centerline{\includegraphics[width=10cm]{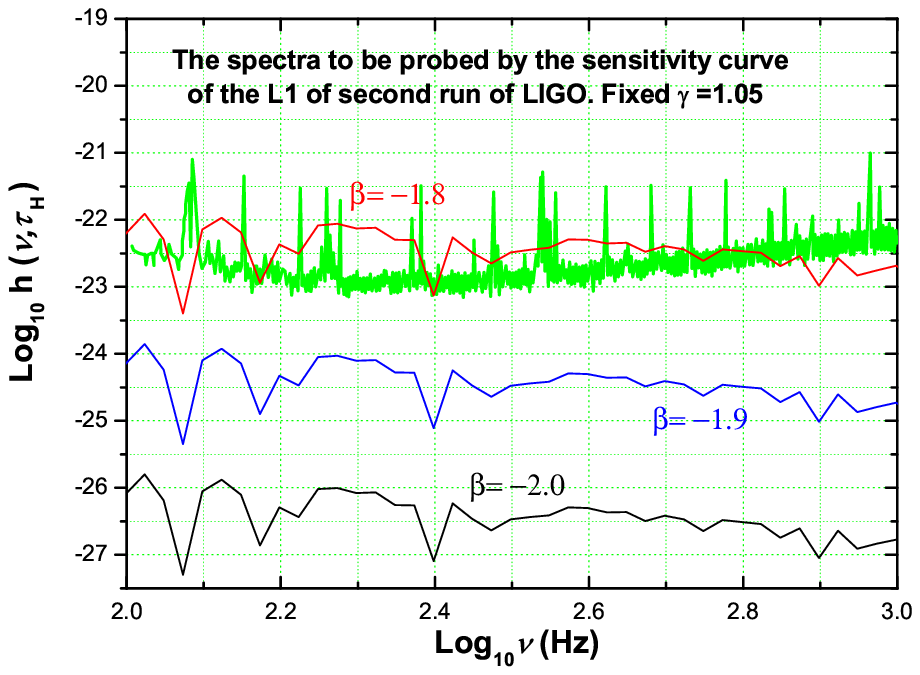}}
\caption{\label{ligo-105} For a fixed acceleration parameter
$\gamma=1.05$ the exact spectrum $h(\nu,\tau_H)$ is plotted within
the range of $\nu= 10^2-10^3$Hz for three inflationary models of
$\beta=-1.8$, $\beta=-1.9$, and $\beta=-2.0$, to compare  with the
sensitivity curve of second run from LIGO L1 \cite{b-abbott2}.}
\end{figure}

\begin{figure}
\centerline{\includegraphics[width=10cm]{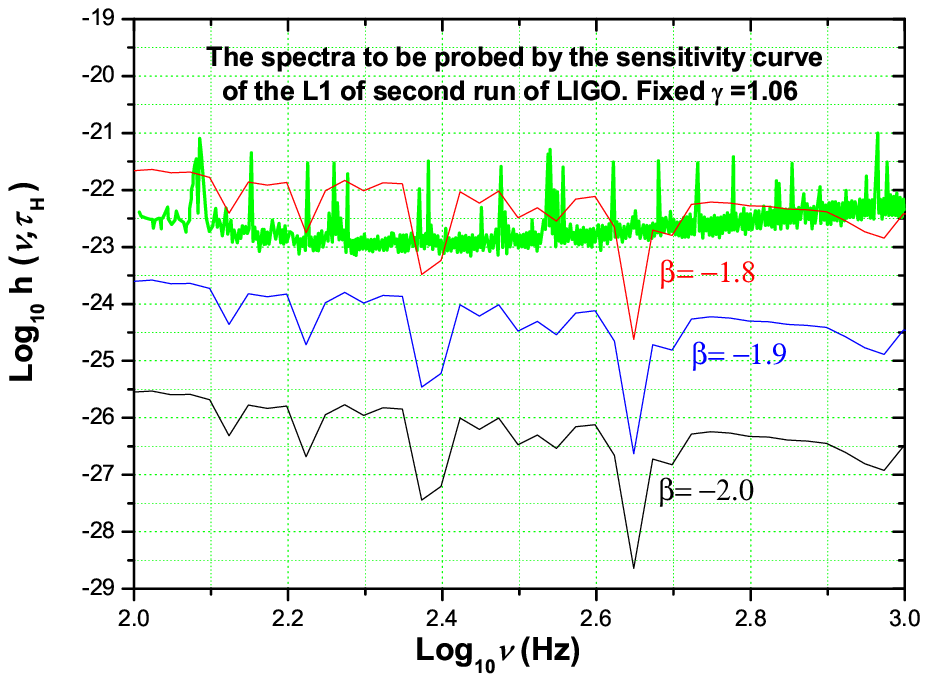}}
\caption{\label{ligo-106}For a fixed acceleration parameter
$\gamma=1.06$ the exact spectrum $h(\nu,\tau_H)$ is plotted within
the range of $\nu= 10^2-10^3$Hz for three inflationary models of
$\beta=-1.8$, $\beta=-1.9$, and $\beta=-2.0$, to compare  with the
sensitivity curve of second run from LIGO L1 \cite{b-abbott2}.}
\end{figure}

Now we make a comparison of the exact spectrum $h(\nu, \tau_H)$
with the sensitivity curve from
the recent S2 of LIGO  \cite{abbott} \cite{r-abbott} \cite{b-abbott2}
with the sensitivity $10^{-22}$ to $ 10^{-23}$
in the  frequency range  $\nu = 10^2\sim 10^3$Hz.
$h(\nu, \tau_H)$ is given in Fig.\ref{ligo-105} for $\gamma=1.05$
and in Fig.\ref{ligo-106} for $\gamma=1.06$.
Both figures have plotted
three spectra for inflationary models $\beta=-1.8$, $\beta=-1.9$, and $\beta=-2.0$,
respectively.
It is found  that the  inflationary models with $\beta \geq -1.8$ has
an amplitude about an order higher than the LIGO sensitive curve.
Even if we take a much lower  value for the (tensor/scalar) ratio,
say $r=0.05$,
the spectrum is still within the region detectable by the LIGO.
Thus,
the inflationary model $\beta = -1.8$  generating the relic GW with $r>0.05$
is ruled out by the LIGO null results.
The models $\beta \leq -1.9$ are still alive by this test alone.
Moreover, when LIGO reaches its design sensitivity $ \sim 10^{-24}$
in the frequency range
in the forthcoming runs,
it will also be able to test the model of $\beta=-1.9$.

\begin{figure}
\centerline{\includegraphics[width=10cm]{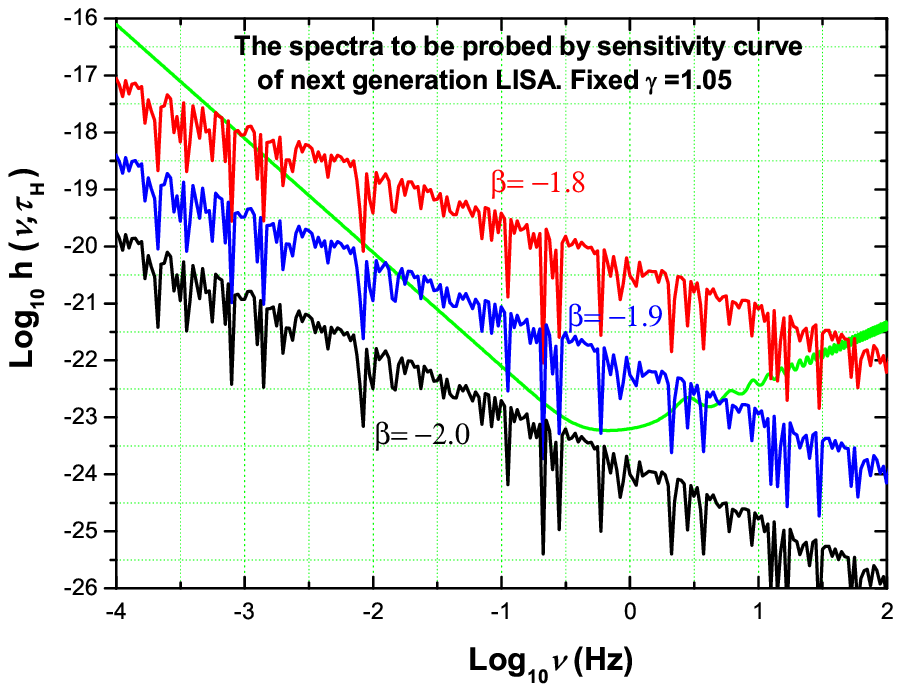}}
\caption{\label{lisa-105} For a fixed acceleration parameter
$\gamma=1.05$ the exact spectrum $h(\nu,\tau_H)$ is plotted within
the range of $\nu= 10^{-4}-10^2$Hz for three inflationary models
of $\beta=-1.8$, $\beta=-1.9$, and $\beta=-2.0$, to compare  with
the sensitivity from LISA the Next Generation \cite{lisa}.}
\end{figure}

\begin{figure}
\centerline{\includegraphics[width=10cm]{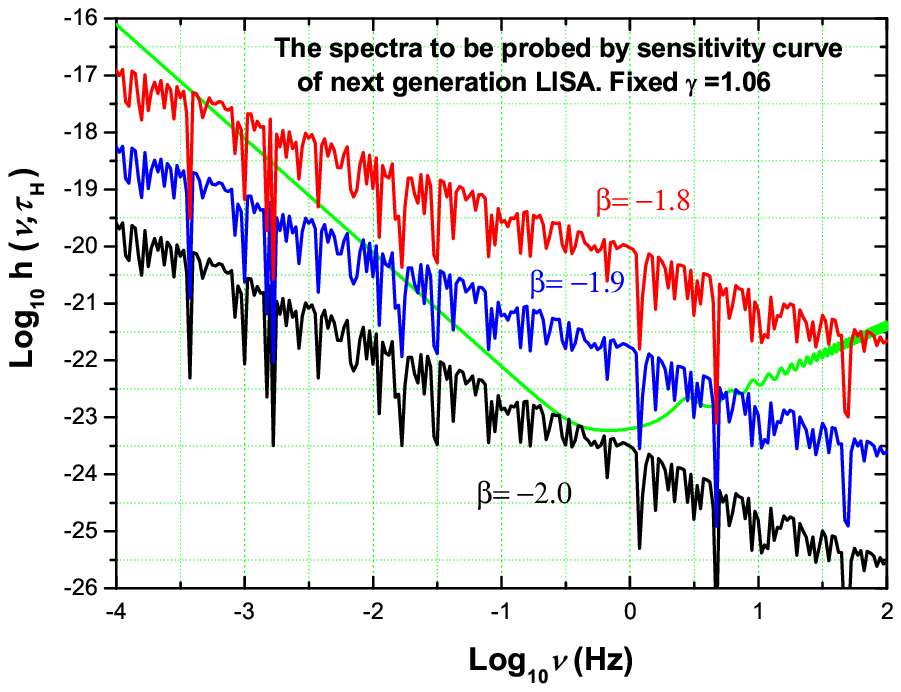}}
\caption{\label{lisa-106}For a fixed acceleration parameter
$\gamma=1.06$ the exact spectrum $h(\nu,\tau_H)$ is plotted within
the range of $\nu= 10^{-4}-10^2$Hz for three inflationary models
of $\beta=-1.8$, $\beta=-1.9$, and $\beta=-2.0$, to compare  with
the sensitivity from  LISA the Next Generation \cite{lisa}.}
\end{figure}

Fig.\ref{lisa-105} for $\gamma=1.05$ and Fig.\ref{lisa-106} for $\gamma=1.06$
give a comparison of the exact spectrum $h(\nu, \tau_H)$
with the sensitivity curve from LISA the Next Generation \cite{lisa}
in the lower frequency range
$\nu = 10^{-4} \sim 10^2$Hz.
It is interesting to notice that,
when the LISA, as being designed, runs in space in the near future,
it will be able to examine directly not only the model $\beta=-1.8$
but also the model $\beta =-1.9$.
For the latter model,
even if a much lower value of the ratio $r=0.05$ is taken,
the LISA will still be able to detect it.
This will be an improvement over the LIGO detection on the earth.
However, as the two figures  show,
the inflationary model $\beta =-2.0$ seems to be still difficult
to detect by the LISA with the capability as presently designed.

\begin{figure}
\centerline{\includegraphics[width=10cm]{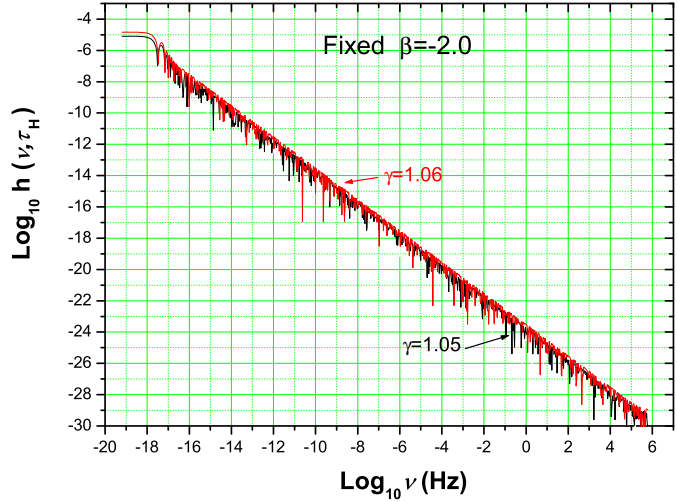}}
\caption{\label{2-0amplitude} For a fixed inflationary parameter
$\beta=-2.0$
 the spectrum $h(\nu,\tau_H)$ is plotted
for different acceleration models of $\gamma=1.05$ and $\gamma=1.06$.
The two spectra are quite close to each other,
and the difference in amplitudes of $h(\nu,\tau_H)$ is quite small, and
difficult to tell in this figure.}
\end{figure}

\begin{figure}
\centerline{\includegraphics[width=10cm]{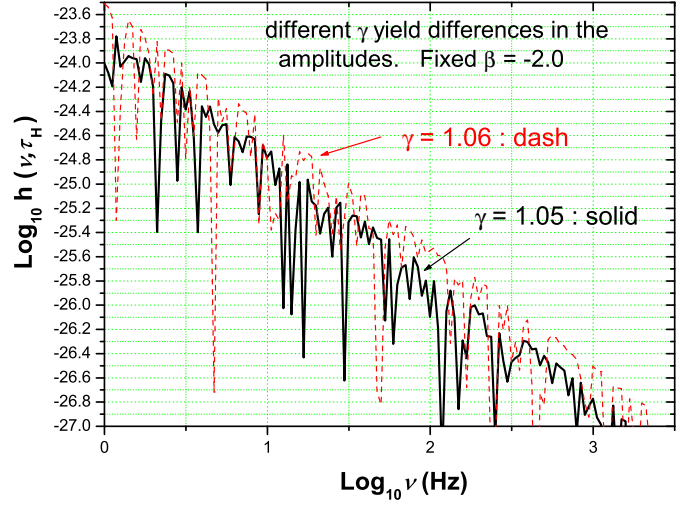}}
\caption{\label{fine20} This enlarged picture is a portion of
Fig.\ref{2-0amplitude} in the range $\nu =1-10^3$ Hz to show fine
differences in the spectrum $h(\nu,\tau_H)$ for different
acceleration models. Note that the amplitude of $h(\nu,\tau_H)$
for the model $\gamma=1.06$ is about $\sim 50 \%$ higher than that
of model $\gamma=1.05$. But in the range $\nu = 10^2-10^3$Hz the
amplitude is only about $\leq 3\times 10^{-26}$, not accessible to
the current LIGO yet.}
\end{figure}

\begin{figure}
\centerline{\includegraphics[width=10cm]{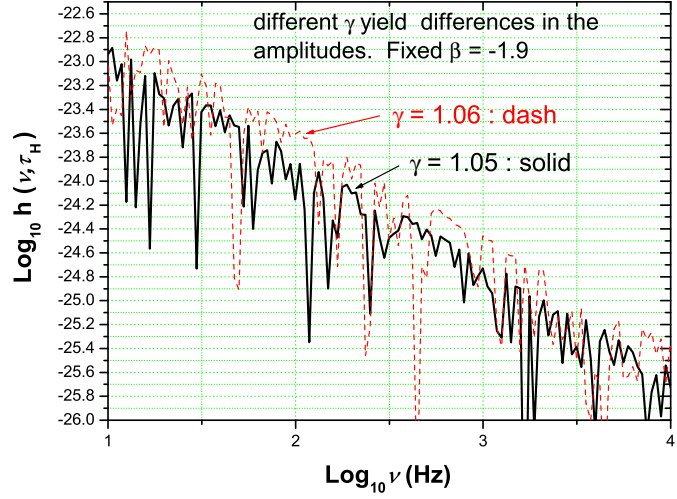}}
\caption{\label{fine19} For a fixed  $\beta=-1.9$ this enlarged
picture in the range $\nu =10-10^4$Hz shows fine differences in
the spectrum $h(\nu,\tau_H)$ for different acceleration models.
Again the amplitude of $h(\nu,\tau_H)$ for the model $\gamma=1.06$
is about $\sim 50 \%$ higher than that of model $\gamma=1.05$. Now
in the range $\nu = 10^2-3\times 10^2$Hz the amplitude is about
$\sim 10^{-24}$, accessible to the LIGO as it approaches its
designed sensitivity $10^{-24}$.}
\end{figure}

Let us examine the dependence of the spectrum $h(\nu, \tau_H)$
upon the dark energy $\Omega_{\Lambda}$ through
the acceleration model parameter $\gamma$.
In Fig.\ref{2-0amplitude} for a fixed $\beta=-2.0$ we plot two
spectra $h(\nu, \tau_H)$ for the acceleration models $\gamma=1.05$ and $\gamma=1.06$
in a broad range of frequencies.
As is seen, the difference between
these two acceleration models are small.
To show the details in the enlarged pictures,
in Fig.\ref{fine19} and Fig.\ref{fine20}
we have  plotted the spectra in a narrow range of frequencies.
It can be read that the amplitude of the model $\gamma=1.06$ is about
$\sim 50 \%$ greater than that of the model $\gamma=1.05$.
That is, in the accelerating Universe with $\Omega_{\Lambda} =0.65$
the amplitude of relic GW is $\sim 50 \%$ higher than
the one with $\Omega_{\Lambda} =0.7$.
Note that the spectrum amplitude $h(\nu, \tau_H)$ itself is very small,
so this amount of $\sim 50 \%$ of difference is probably difficult
to detect at present.
However, in principle, it does provide
a new way to tell the dark energy fraction $\Omega_{\Lambda}$  in the Universe.
With the LIGO approaching its designed sensitivity,
hopefully this difference can be detected thereby.
As the LISA is currently designed, it will also be able detect this effect.

\begin{figure}
\centerline{\includegraphics[width=10cm]{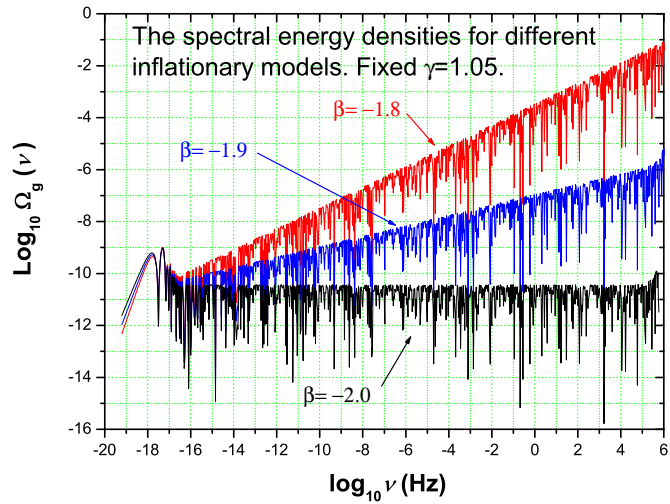}}
\caption{\label{energy105} For a fixed $\gamma=1.05$ the spectral
energy density $\Omega_g(\nu)$ is plotted for the models of
$\beta=-1.8$, $\beta=-1.9$, and $\beta=-2.0$. Obviously, the
inflationary model of $\beta=-1.8$ has an $\Omega_g(\nu)$
increasing too fast with the frequency $\nu$, thus is ruled out by
the LIGO bound and the necleosynthesis bound. $\Omega_g(\nu)$ in
the model of $\beta=-1.9$ is narrowly below the necleosynthesis
bound, but since $\Omega_g(\nu)$ increases also too fast with
$\nu$ so it will barely survive. The model of $\beta=-2.0$ has a
flat spectral energy density with a value $\sim 10^{-10}$, much
smaller than the necleosynthesis bound. Thus the model
$\beta=-2.0$ is robust.}
\end{figure}

\begin{figure}
\centerline{\includegraphics[width=10cm]{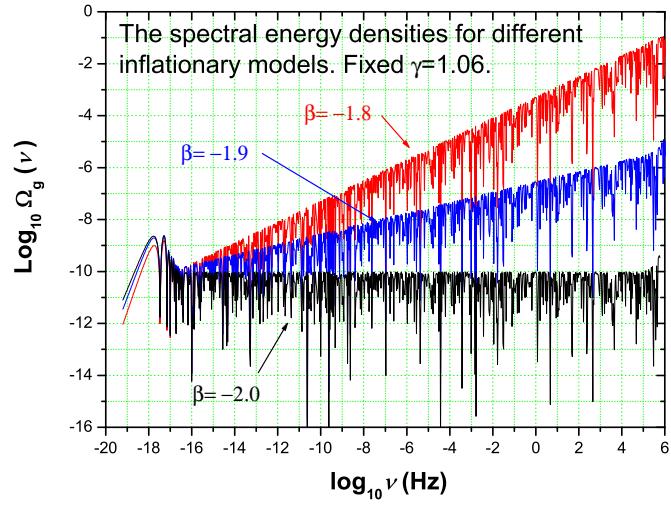}}
\caption{\label{energy106} This picture is similar to
Fig.\ref{energy105} but for a fixed $\gamma=1.06$. The conclusions
are also similar to Fig.\ref{energy105}.}
\end{figure}

Let us examine the spectral energy density $\Omega_g(\nu)$ and their constraints.
Fig.\ref{energy105} and Fig.\ref{energy106}
are the plots of the spectral energy density
$\Omega_g(\nu)$ defined in Eq.(\ref{omega}) for $\gamma=1.05$ and $\gamma=1.06$,
respectively.
These plots of the exact analytic results agree
with the numerical one in \cite{zh2}.
If we use the result LIGO third science run \cite{b-abbott1}
of the energy density bound for the flat spectrum with
$\Omega_0 <8.4\times 10^{-4}$ in the $69-156$ Hz band,
then the model $\beta=-1.8$ is ruled out, but the models $\beta \leq-1.9$ survive.
However, this LIGO constraint on the GW energy density  is not as stringent as
the constraint by
the so-called nucleosynthesis bound \cite{maggiore} \cite{maia2},
whose main idea is the following:
In the early Universe at a temperature $T\sim $ a few  Mev
the nucleosynthesis process is going on.
The relic GW will contribute to the total energy density $\rho$
that drives the Universe expansion,
thus will increase the effective number of species of particles $g_*$.
More relic GW energy will enhance the freeze-out temperature for the process
$pe\leftrightarrow n\nu$, and will lead to more neutrons available for
the production of helium-4 ($^4 He$).
In practice the effective number of neutrino species $N_{\nu}$
is used in place of $g_*$.
Analysis has led to the nucleosynthesis bound on the relic GW energy density
at the present time \cite{maggiore}:
\be \label{energylimit}
\int \Omega_g(\nu) \, d(\log \nu) \leq  0.56\times 10^{-5}.
\ee
where the value $\rho_{\gamma} \simeq 2.481\times 10^{-5} \rho_c$
and the conservative value $N_{\nu}< 4$ have been used.
Note that this is bound on the total GW energy density integrated over
all frequencies.
The integrand function should also have a bound
$ \Omega_g(\nu) < 0.56\times 10^{-5}$
in the interval of frequencies $\delta (\log \nu) \simeq 1$.
By this constraint it is also seen from Fig.\ref{energy105} and Fig.\ref{energy106}
 that the model $\beta =-1.8$ has an
$\Omega_g(\nu)$ too high, is therefore ruled out,
the same conclusion that we arrived at from Fig.\ref{ligo-105} and Fig.\ref{ligo-106}.
The model with $\beta = -1.9$ are barely alive,
as its energy density $\Omega_g(\nu)$ tends to be growing higher
with  high frequencies.
The model $\beta = -2.0$ are still robust since its spectral energy density
$\Omega_g(\nu)$ is
a flat function much lower than the limit
 in Eq.(\ref{energylimit}).


\section{Analytic Approximation}

We now want to give an approximation to the above exact solution $
h(k,\tau_H)$ to recover the approximate analytic one given in
\cite{zh}. The following approximation for the Bessel functions
will be used
 \be \label{j}
 J_\mu(x)\approx  \sqrt {\frac{2}{\pi x}}
 \cos(x-\frac{\mu  \pi}{2}-\frac{\pi}{4}), \,\,\,\,\,
  x\gg 1,
 \ee
 \be J_\mu(x)\approx \frac{1}{\Gamma(\mu+1)}(\frac{x}{2})^\mu, \,\,\,\,\,
  x\ll 1.
 \ee
Note that the  coefficients $D_1,D_2$, $B_1$, $B_2$, $C_1$, $C_2 $,
$E_1$, and $E_2$ are all functions of $k$,
and they need to be approximated according to the value of $k$.

In the long-wave limit, $\ x_1=k\tau_1 \ll 1$
and $t_1 = (1+\beta_s)x_1/(1+\beta) \ll 1$,
from  Eqs.(\ref{d1}) and (\ref{d2}) one has

\be
D_1\approx  2^{-\beta+\beta_s}   (\frac{1+\beta}{1+\beta_s})^{\beta+1}
            t_1^{\beta-\beta_s}A_1,
\,\,\,\,\,\,\,
D_2\approx
                t_1^{\beta+\beta_s+3}A_1.
\ee
$D_2$ is higher order of $t_1$ and can be neglected in the following.

From Eqs.(\ref{b1}) and (\ref{b2}),
in the long-wave limit  $t_s\ll 1$ and  $y_s\ll 1$, one has
\bee
B_1\approx
 it_s^{\beta_s }D_1 \propto k^\beta,
\,\,\,\,\,\,\,   B_2\approx -B_1.
\een

From Eqs.(\ref{c1}) and (\ref{c2}) , in the long-wave limit $k\ll 1/\tau_2$,
one has
\be
C_1\approx -\frac{3i}{2y_2}B_1\sim k^{\beta-1},
\,\,\,\,\,  C_2 \ll C_1,
\ee
so  $C_2$ can be neglected.
In the shortwave limit $k\gg 1/\tau_2 $, one has
\be
C_1\approx -2iB_1\sin z_2, \,\,\,\,\,
C_2\approx 2iB_1\cos z_2  .
 \ee

From Eqs.(\ref{e1}) and (\ref{e2}), for $k\tau_E\ll 1$,
one has
\be  \label{e}
E_1\approx C_1, \,\,\,\,\, E_2\approx C_2,
\ee
which  also holds approximately for $k\tau_E\gg 1$,
with some extra oscillating factors.

With all these coefficients being estimated,
now we can evaluate the approximation of
the  spectrum in Eq.(\ref{spectrum})
at the present time $\tau_H$, which is written as
\[
  h(k,\tau_H)   = A\frac{l_0}{2\pi   b}k |\frac{\mu_k(\tau_H)}{a(\tau_H)}|.
\]
Substituting the expressions Eq.(\ref{m}) for $\mu_k(\tau_H)$
and   Eq.(\ref{l0}) for $l_0$ into the above  leads to
\be
   h(k,\tau_H)=
   A[\gamma (\zeta_E^{-(1+\frac{1+\beta}{\gamma})}
   \zeta_2^{\frac{\beta-1}{2}}
   \zeta_s^{\beta}
   \zeta_1^{\frac{\beta_s-\beta}{1+\beta_s}}]\frac{k}{k_H}
   \sqrt{\frac{\pi s_H}{2}}
   |E_1J_{\gamma+\frac{1}{2}}(s_H)+E_2J_{-\gamma -\frac{1}{2}}(s_H)|.
\ee
Using the results from Eq.(\ref{j}) through Eq.(\ref{e}),
we approximate this expression by the leading term of power-law of $k$
in various ranges of $k$.
By some straightforward calculations, using
$|(\tau_H-\tau_a)/(\tau_E-\tau_2)|=1/(1+z_E)$,
we obtain the following
expressions for the analytic approximate spectrum
\be \label{ini}
h(k, \tau_H) = A(\frac{k}{k_H})^{2+\beta}, \,\,\,\,\, k\leq k_E;
\ee
\be \label{eh}
h(k,\tau_H)\approx
   A (\frac{k}{k_H})^{\beta-1}\frac{1}{(1+z_E)^{3+\epsilon}},\,\,\,\,\,
   k_E \leq k \leq k_H;
\ee
\be \label{h2}
h(k,\tau_H) \approx
   A      (\frac{k}{k_H})^\beta   \frac{1}{(1+z_E)^{3+\epsilon}},\,\,\,\,\,
   k_H \leq k \leq k_2;
\ee
\be \label{2s}
h(k,\tau_H)
  \approx  A   (\frac{k}{k_H})^{\beta+1}  \frac{k_H}{k_2}
  \frac{1}{(1+z_E)^{3+\epsilon}},
  \,\,\,\,\,\,\, k_2 \leq k \leq k_s;
\ee
\be \label{fin}
  h(k,\tau_H) \approx
 A    (\frac{k_s}{k_H})^{\beta_s}\frac{k_H}{k_2}
    (\frac{k}{k_H})^{\beta-\beta_s+1}\frac{1}{(1+z_E)^{3+\epsilon}},
     \,\,\,\,\,\,\, k_s \leq k \leq k_1,
\ee where the small parameter $\epsilon \equiv
(1+\beta)(1-\gamma)/\gamma$, also depending on the behavior of the
acceleration expansion through $\gamma$. The model $\gamma= 1$
gives $\epsilon =0$, and the results of Eqs.(\ref{ini}) through
(\ref{fin}) reduce to exactly our early result  given in
\cite{zh}.
The influence of detailed accelerating expansion on the
$h(k, \tau_H)$ is mainly demonstrated through the factor
$1/(1+z_E)^{3+\epsilon}$, causing a variation in the magnitude of
$h(k,\tau_H)$.
For the inflationary expansion with $\beta\approx
-2$ , the model of $\gamma =1.05$ ($\Omega_{\Lambda}=0.7$) gives
$1/(1+z_E)^{3+\epsilon}= 0.423$, and the model $\gamma = 1.06$
($\Omega_{\Lambda}=0.65$) gives $1/(1+z_E)^{3+\epsilon}=0.533$,
yielding roughly the amplitude of the model $\gamma=1.06$ greater than
that of the $\gamma=1.05$ by about $\sim 30\%$.
The more accurate computation from the exact solutions shows
an average difference of $\sim 50 \%$,
as plotted in Figs.(\ref{fine19}) and (\ref{fine20}).
Note that the factor $1/(1+z_E)^{\epsilon}=0.987$ for the model $\gamma =1.05$,
and $1/(1+z_E)^{\epsilon}=0.989$ for the model $\gamma =1.06$,
differing by only $0.2 \%$,
too small to tell by the current experimental detections.
Therefore, in regards to the amplitude of relic GW,
one can simply put $\epsilon=0$ in the approximate spectrum given
in Eqs.(\ref{ini}) through (\ref{fin}), just as it was in the model $\gamma=1$,
causing only a difference of $0.2 \%$ in the amplitude for a variety of models
with various $\gamma$.

We remark that each of these expressions
from Eq.(\ref{eh}) to (\ref{fin}) holds up to a numerical factor
$A$, which  contains certain oscillating factors of the form
$\cos(k\tau_H)$, or $\cos(y_2)$ and $\sin(t_s)$.
In comparison
with the decelerating models \cite{grishchuk}, Eq.(\ref{eh}) is a
new segment of spectrum in $k_E<k< k_H$, whose occurrence is due
to the acceleration of current expansion of the Universe. Besides,
the three segments of spectrum, i.e.,
 Eqs .(\ref{h2}), (\ref{2s}),  and (\ref{fin}),
all have the extra factor
$(1+z_E)^{-3-\epsilon} = (\Omega_m/ \Omega_{\Lambda})^{1+\epsilon/3}$
that are missing in the corresponding three segments
in the decelerating models.


\section{Conclusion}

We have presented a detailed calculation of the exact analytic spectrum of
relic GW in the present flat $\Omega_{\Lambda}+\Omega_m = 1$
Universe in accelerating expansion.
The resulting exact spectrum explicitly depends on
the detailed behavior of the present accelerating expansion,
characterized by the parameter $\gamma$
in the scale factor $a(\tau)\propto |\tau|^{-\gamma}$.
It also explicitly depends on the inflationary model $\beta$,
the reheating model $\beta_s$, and the  (tensor/scalar) ratio $r$ as well.
Therefore, the result is general enough to
describe the GW spectrum $h(\nu,\tau_H)$
produced from in a variety of accelerating cosmological models.
One can use the formula in other applications
by choosing a set of parameters $\beta$, $\beta_s$, $\gamma$, and $r$.
Besides, the analysis of the exact result gives the following conclusions:

The GW amplitude of the model $\gamma=1.06$ is about
$\sim 50 \%$ greater than that of the model $\gamma=1.05$,
i.e., in the accelerating Universe with $\Omega_{\Lambda} =0.65$
the amplitude of relic GW is $\sim 50 \%$ higher than
the one with $\Omega_{\Lambda} =0.7$.
Although it is probably difficult to detect at present,
the effect does provide a new way to tell
the dark energy fraction $\Omega_{\Lambda}$  in the Universe.
Hopefully this difference can be detected
when the LIGO approaches its designed sensitivity   $ \sim 10^{-24}$,
and the LISA runs in future.

The  spectrum  depends sensitively on the parameter $\beta$
of the inflationary models.
A larger value of $\beta$ yields  a flatter spectrum
$h(\nu,\tau_H)$ with more power on the higher frequencies.
The sensitivity curve of current LIGO
rules out the inflationary models with  $\beta\geq-1.8$.
The LIGO  with its design sensitivity
and the LISA in future will also be able to test the  $\beta=-1.9$ model  directly.

The relic GW is also constrained through
its spectral energy density $\Omega_g(\nu)$ by
the resent LIGO bound and the nucleosynthesis bound.
While both bounds rule out the inflationary model $\beta=-1.8$,
the nucleosynthesis bound puts the model $\beta=-1.9$ in danger.
However, the model $\beta=-2.0$ (de Sitter) is robust,
since its spectral  energy density $\Omega_g(\nu)$ is flat and is
$\sim 10^{-10}$, much smaller than the nucleosynthesis bound.

Finally, the exact analytic spectrum reduces to
the approximate analytic and the numerical ones given  in our previous study
for the case $\gamma=1$.


\section*{Acknowledgements}
Y. Zhang's research work has been supported by the Chinese NSF (10173008),
NKBRSF (G19990754), and by SRFDP.


\newpage

\end{document}